\documentclass[prx,twocolumn,aps,epsf,showpacs,superscriptaddress,longbibliography,nofootinbib]{revtex4-1}
\usepackage[pdftex]{graphicx}
\usepackage[normalem]{ulem}
\usepackage{verbatim, float, complexity}
\usepackage{dcolumn, amsfonts, mathtools}
\usepackage{bm}
\usepackage{latexsym} 
\usepackage{amsmath}
\usepackage{dsfont}
\usepackage{MnSymbol}
\usepackage[dvipsnames]{xcolor}
\usepackage{array}
\usepackage{bbm}
\usepackage{color}
\usepackage{array}
\usepackage{cancel}
\usepackage{braket}
\usepackage{comment}
\usepackage{overpic}
\DeclareUnicodeCharacter{2013}{--}
\usepackage[colorlinks=true,allcolors=blue]{hyperref}%

\newcommand{\<}{\langle}

\renewcommand{\>}{\rangle}
\renewcommand{\(}{\left(}
\renewcommand{\)}{\right)}
\renewcommand{\[}{\left[}
\renewcommand{\]}{\right]}

\begin{document}

\title{Boundary Criticality in the 2d Random Quantum Ising Model}
\author{Gaurav Tenkila}
\affiliation{Department of Physics and Astronomy, and Quantum Matter Institute, University of British Columbia, Vancouver, BC, Canada V6T 1Z1}

\author{Romain Vasseur}
\affiliation{Department of Physics, University of Massachusetts, Amherst, MA 01003, USA}

\author{Andrew C. Potter}
\affiliation{Department of Physics and Astronomy, and Quantum Matter Institute, University of British Columbia, Vancouver, BC, Canada V6T 1Z1}

\begin{abstract}
The edge of a quantum critical system can exhibit multiple distinct types of boundary criticality.
We use a numerical real-space renormalization group (RSRG) to study the boundary criticality of a 2d quantum Ising model with random exchange couplings and transverse fields, whose bulk exhibits an infinite randomness critical point.
This approach enables an asymptotically numerically exact extraction of universal scaling data from very large systems with many thousands of spins that cannot be efficiently simulated directly.
We identify three distinct classes of boundary criticality, and extract key scaling exponents governing boundary-boundary and boundary-bulk correlations and dynamics.
We anticipate that this approach can be generalized to studying a broad class of (disordered) boundary criticality, including symmetry-enriched criticality and edge modes of gapless symmetry-protected topological states, in contexts were other numerical methods are restricted to one-dimensional chains.
\end{abstract}

\maketitle

%\tableofcontents

The edge of a critical system whose bulk is poised at a phase transition, can exhibit multiple different classes of types of boundary criticality with distinct universal scaling properties~\cite{diehl1989semiinfinite, deng2005surface, deng2006bulk,ohno19841, landau1989monte}. 
For example, a (quantum) Ising model in two or spatial dimensions has an extensive boundary that is capable of supporting its own long-range magnetic order independent from the bulk. This leads to multiple distinct classes of boundary criticality depending, for example, on whether or not the boundary has long-range order.
While the study of boundary criticality has a long history~\cite{diehl1997thetheory, cardy1996scaling}, there has been a recent resurgence of interest in exploring boundary criticality near the lower-critical dimension~\cite{metlitski2022on, sun2022quantum}, the influence of symmetry-enriched criticality and topological edge states on boundary \emph{quantum} criticality~\cite{weber2018nonordinary}, and boundary-distinctions between Higgs and confined phases of gauge theories~\cite{verresen2022higgs}. 
Analyzing boundary criticality in quantum many-body settings more than one spatial dimension presents challenging for both analytic and numerical schemes. The inherent mixed-dimensionality that makes it challenging to employ standard analytic schemes such as dimensional regularization. Moreover, in numerics, strong finite size effects also make it hard to cleanly disentangle bulk- and boundary- criticality.

In this work, we adapt the strong-disorder real-space renormalization group (RG) approach~\cite{ma1980lowtemperature,fisher1992random,fisher1995critical,motrunich2000infinite} to study boundary criticality in disordered quantum systems, demonstrating this approach to map out the boundary phase diagram and critical scaling properties of a 2d Ferromagnetic Ising model with random- exchange interactions and transverse fields.
In addition to possible relevance to disordered quantum magnetic materials, this approach provides a rare example of asymptotically exact technique for extracting boundary criticality in a 2d quantum system. 
As with other low-dimensional strong-disorder RG approaches, the approximations of this technique are controlled by a flow to asymptotically-infinite randomness.
While strong-disorder RG schemes can be analyzed analytically in $1d$, numerical implementation appears necessary in $2d$, with only recent developments addressing this problem analytically~\cite{pandey2023random}. However, unlike direct simulation of the quantum many-body problem, whose complexity scales exponentially in the number of spins, the numerical RG steps can be efficiently implemented on extremely large system sizes, enabling accurate determination of critical properties.

\begin{figure}[t]
\includegraphics{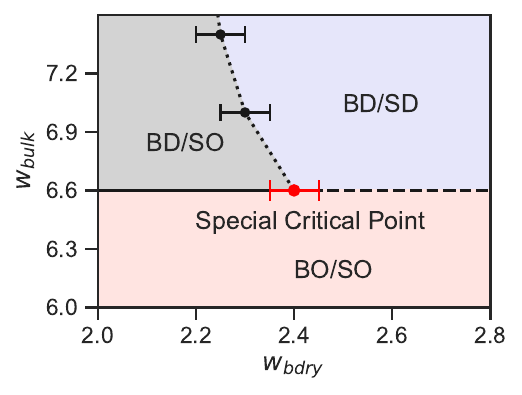}
\caption{{\bf Bulk and Boundary phase diagram of the $2d$ random Ising model.} $w_{\rm bulk},w_{\rm bdry}$ are parameters controlling the widths of the distribution of random ferromagnetic Ising couplings in the bulk and boundary (for fixed distribution of transverse fields). Phases are indicated by letters: $B$ for bulk or $S$ for surface/boundary, and $D$, $O$ for disordered or ordered respectively.}
\label{fig:phase_diag}
\end{figure}

Here, we use the strong-disorder RG method to study the boundary phase diagram of a 2d random bond- and transverse-field- Ising model, whose bulk is fixed at a (strong-randomness) quantum phase transition~\cite{motrunich2000infinite, kovacs2010renormalization,yu2008entanglement,kovacs2013boundary} between ferromagnet and paramagnetic phases.
By tuning the competition of the exchange interactions and transverse fields on the boundary, we identify three distinct infinite-randomness boundary quantum critical (BQC) classes that are infinite-randomness analogs of those identified in clean Ising models. These include: i) an \emph{ordinary} BQC phase in which the transverse fields dominate at the boundary and spin correlations exhibit strong-disorder critical random-singlet like scaling, ii) an \emph{extraordinary} BQC phase in which the random exchange couplings dominate and the edge exhibits long-range magnetic order, which induces a fractal structured magnetization profile that slowly decays as one moves away from the edge, and iii) a \emph{special} BQC marking the boundary phase transition between ordinary and extraordinary.
Each of these universality classes exhibits strong randomness behavior characterized by a distinct stretched-exponential decay of typical correlation functions and tunneling rates with distance, and power-law decay of average correlation functions. Further, the critical surface magnetization for the ordinary transition is in good agreement with previous numerical work done in this regime~\cite{kovacs2013boundary}.

We expect that these methods can be adapted to a wide variety of disordered models with discrete symmetry, including those where symmetry and topology enrich the bulk criticality~\cite{scaffidi2017gapless,verresen2021gapless}.

\section{Model and Strong-Disorder RG}
Following~\cite{motrunich2000infinite}, we consider a random $2d$ transverse-field Ising model on a triangular lattice:
\begin{equation}
H = -\sum_i h_i \sigma_i^x -\sum_{\langle ij\rangle} J_{ij} \sigma_i^z \sigma_j^z,
\label{eq:hising}
\end{equation}
where $\sigma^{x,y,z}$ are spin-1/2 Pauli matrices, and the transverse fields $h_i$ and ferromagnetic Ising couplings (``bonds") $J_{ij}$ are drawn from probability distributions which we describe below . The distributions are each characterize by a single parameter, $w$, that controls the variance of the logarithm of these coupling constants. We will eventually tune this parameter independently for boundary and bulk couplings.

The strong disorder RG scheme~\cite{ma1980lowtemperature,fisher1992random} proceeds by identifying the strongest coupling, projecting the system into the ground-state of this coupling, and perturbatively computing the residual couplings between the spins once the strong coupling is eliminated.

When the strongest coupling is a ferromagnetic bond, $J_{ij}$, the two spins, $i,j$, involved are locked into one of two ferromagnetically-aligned configurations and reduce to a single cluster with two states, that experiences a much-reduced transverse field (describing quantum tunneling between the two FM states of the cluster), which to lowest order in perturbation theory is: $\tilde{h} \sim h_ih_j/J_{ij}$.
Here, the proportionality constant is irrelevant for the asymptotic scaling behavior. 
As the RG proceeds, further decimations of strong bonds can create ever larger clusters. 
At criticality, fractal-geometry clusters emerge, with scale-invariant distribution of sizes.

When the strongest coupling is a transverse field on spin (or more generally, a ferromagnetic cluster) $i$, the spin/cluster is locked to the local field, and drops out of the future rounds of the RG. New exchange interactions are generated by virtual super-exchange processes to all spins $j,k$ coupled to $i$: with strength $\tilde{J}_{jk} \sim J_{ij}J_{ik}/h_i$.
The RG scheme is then iterated, proceeding to the next strongest coupling.
In the process, the distribution of exchange and transverse field couplings, as well as the graph of interactions evolves (``flows").
However, no new types of couplings besides random fields and Ising interactions are generated.

%The strong disorder RG scheme proceeds by identifying the strongest local term in (\ref{eq:hising}), and treats the other competing couplings peturbatively. 
%When the strongest term is a random transverse field on site $i$, to lowest order, the field pins $\sigma^x_i=1$. 
%Spin
%%Terms that do not commute with the strongest one are eliminated perturbatively perturbatively, yielding the following effective Hamiltonian terms:
%\begin{align}
%\text{Ising Bond decimation: } &\frac{h_i h_j}{J_{ij}} \sigma_i^x \sigma_j^x \\
%\text{Transverse Field Decimation: } &\frac{J_{ki} J_{il}}{h_i} \sigma_k^z \sigma_l^z
%\end{align}

%This defines an RG flow on the probability distribution of the couplings. Conveniently, no new types of interactions besides random  form of the  since the new terms generated serve to only modifying the coupling constants of the original Hamiltonian. 

%Since the RG decimation rules for the couplings are multiplicative in the couplings, it is convenient to recast them in terms of their logarithms variables makes the rules additive, considerably simplifying the numerics. To this extent, we employ the same convention used by Motrunich et al.

Along the RG flow, the width of these distributions grows asymptotically without bound, i.e. the flow is to infinite randomness, and a fixed point distributions are only observed in appropriately RG-scale dependent logarithmic variables:
\begin{align}
    \beta_i &= \log (\Omega/h_i),
    ~~~~h_i = \Omega e^{-\beta_i}
    \nonumber \\
    \zeta_{ij} &= \log ( \Omega/J_{ij}),
    ~~~~J_{ij} = \Omega e^{-\zeta_{ij}} 
\end{align}
Here, $\Omega$ denotes the scale of the strongest undecimated couplings at the current RG step. It is also convenient to parameterize the RG scale in terms of its logarithm:
\begin{align}
\Gamma = \log (\Omega_0/\Omega), ~~~~ \Omega = \Omega_0 e^{-\Gamma},
\end{align}
where $\Omega_0$ is the strongest coupling before any RG decimations are carried out.

Refs.~\cite{motrunich2000infinite,kovacs2010renormalization,yu2008entanglement} carried out extensive numerical simulations of the bulk criticality of a 2d Ising model. In the following sections we briefly review key findings, which we utilize below to study boundary criticality. 

\subsection{Coupling distributions}
Ref.~\cite{motrunich2000infinite} observed that the distribution of transverse fields flows to an exponential distribution in the log variables defined above:
\begin{align}
    R(\beta,\Gamma) &=R_0(\Gamma)e^{-R_0(\Gamma) \beta},
    \label{eq:fielddist}
\end{align}
where $R_0(\Gamma)$ is a numerically-determined function of RG scale $\Gamma$.

In $1d$ random Ising models~\cite{fisher1992random} the same distribution of fields with a simple linear relation $R_0(\Gamma)\sim 1/\Gamma$ holds, and duality between random fields and random bonds dictates an identical form for the random field distribution.
However, in $2d$, the bonds and fields behave fundamentally differently. Instead, the bond distribution was found to be well fit by a linear scaling form:
\begin{align}
    P(\zeta,\Gamma) = a(\Gamma)+b(\Gamma)\zeta,
    \label{eq:bonddist}
\end{align}
where $a(\Gamma),b(\Gamma)$ are numerically determined functions.

To reduce finite-size corrections to scaling, it is useful to sample the initial couplings in (\ref{eq:hising}) from distributions of this form (though neglecting any statistical correlations between the bond and field distributions).
Specifically, the transverse fields, $h_i$, are drawn from (\ref{eq:fielddist}) with $R_0(\Gamma=0)=1$. The bonds are sampled from a log-linear distribution (\ref{eq:bonddist}) with an upper cutoff to $\zeta$ (which translates to a small-$J$ cutoff), $w$:
\begin{align}
    P(\zeta,\Gamma=0)\equiv 
    \begin{cases}
        a+b\zeta & 0\leq \zeta \leq w \\
        0 & {\rm otherwise}.
    \end{cases}
\end{align}
subject to the normalization condition $\int_0^w(a+b\zeta)d\zeta=1$~\footnote{We note that this convention departs from that of Ref.~\cite{motrunich2000infinite}, which used an un-normalized bond distribution. The critical slope and intercept parameters correspond to $a=0.1$ and $w=6.67$.}
This is accomplished by identifying the nearest neighbours for each lattice site and filling the corresponding matrix elements with samples from the probability distribution.
%(generated by the ContinuousRV method in sympy). 
Next-nearest-neighbour couplings can be included in an analogous manner by modifying the bounds of distribution function to be from $w$ to $w'$.

Since only a single tuning parameter is required to hit the transition, we fix $a=0.1$, which matches the value of $a(\Gamma\rightarrow \infty)$ numerically observed at criticality, and tune $w$ ($b$ is then fixed by the normalization requirement).

\subsection{Locating the bulk critical point}
\begin{figure}[t]
{\Large $(a)$}\includegraphics[width=0.45\textwidth]{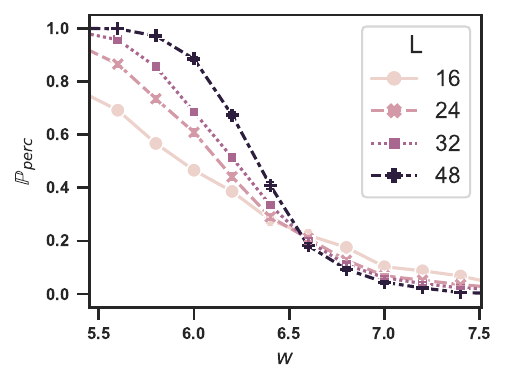}
{\Large $(b)$}\includegraphics[width=0.45\textwidth]{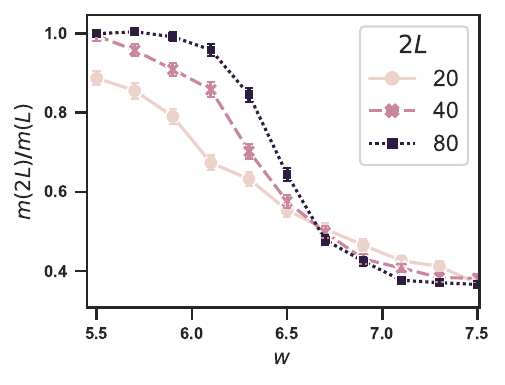}
\caption{{\bf Bulk criticality.}
(a) Probability, $\mathbb{P}_{\rm perc}$, for a cluster to percolate in an $L \times L$ size system with periodic boundary conditions. A cluster is marked as percolating if it connects two ``end-zone" strips defined by $y\in [0,W]$ and $y\in [L/2,L/2+W]$, where we choose the strip width $W=L/8$. 
(b) Ratio of magnetization for linear system sizes L and 2L. 
For both quantities, the crossing of curves for different $L$ at $w=w_c\approx 6.61\pm0.14$ locates the bulk critical point. The data is obtained by averaging over 1440 disorder realizations.
}
\label{fig:perc_PBC}
\end{figure}

To locate the bulk critical point in this model, we follow~\cite{yu2008entanglement}, and examine two quantities that exhibit a stable finite-size crossing that enables one to accurately determine the critical value of the bond-width parameter, $w_c$:

The first quantity we use to locate the bulk transition is the probability, $\mathbb{P}_{\text{perc}}$, to find a  magnetic cluster that percolates across the system. For infinite system size, $L\rightarrow \infty$, $\mathbb{P}_{\text{perc}}$  tends from $1$ deep in the ferromagnetic phase (small $w$) to $0$ deep in the paramagnetic phase (large $w$). In finite size, the crossover between these extreme values is smooth, but the $\mathbb{P}_{\text{perc}}$ curves for increasing sequence of $L$ cross each other at a value of $w$ that quickly converges to $w_c$ as $L\rightarrow \infty$ (Fig.~\ref{fig:perc_PBC}).

We benchmark the location of the critical coupling, $w_c$, identified by the crossing in $\mathbbm{P}_{\rm perc}$ against a second quantity previously used to locate the bulk transition in Ref.~\cite{yu2008entanglement}: the ratio of magnetizations in a system of size $L$ and $2L$: $m(2L)/m(L)$. Numerically, we determine $m$ by running the RG procedure until only a single cluster remains, and taking $m$ to be the fraction of spins in this final cluster (all other spin-clusters are treated as perfectly pinned to their local transverse fields).
Deep in the ferromagnetic (paramagnetic) phase, the ratio $m(2L)/m(L)$ tends to $1$ ($0$) respectively with corrections exponentially-suppressed in the ratio of $L$ to the correlation length $\xi$.
At criticality, the ratio tends to a constant:
\begin{align} \label{eqRatiom}
    \frac{m(2L)}{m(L)} =2^{-x_m},
\end{align}
where $x_m\approx 1.0$ is the anomalous magnetization scaling dimension~\cite{yu2008entanglement}.

\subsection{Correlations and dynamics}
The strong-disorder critical points are characterized by clusters of ferromagnetically-aligned spins with a broad distribution of sizes, and complicated fractal geometries.
In particular, this leads to strong departures in the behavior of average and typical behaviors, which show different sensitivity to rare configurations. 
We briefly review some key features of the resulting physics of correlation functions and dynamics that result from this structure below, and indicate how related universal critical exponents can be extracted (for bulk systems with periodic boundary conditions) from the strong-disorder RG procedure.
In subsequent sections, we show how these methods can be adapted to study boundary criticality in open geometries.

\paragraph{Average correlation functions}
In the strong-disorder RG, the average spin correlation:
\begin{align}
    \overline{G}(r,r') = \overline{\<\sigma^z_r\sigma^z_{r'}\>},
\end{align}
where $\overline{(\dots)}$ denotes averaging over disorder configurations,
can be approximately computed by taking $\<\sigma^z_{r}\sigma^z_{r'}\> \approx  \begin{cases} 1 & r,r' \in \{{\rm same~cluster}\} \\ 0 & {\rm otherwise} \end{cases}$. This is clearly an oversimplification: it neglects reduction of spin correlations within a cluster by quantum fluctuations as well as inter-cluster correlations that decay stretched-exponentially in distance. However, the assumption is that rare configurations where sites $r$ and $r'$ belong to the same ferromagnetic cluster dominate the average correlation function, and capture the correct asymptotic universal scaling at long distances.

By contrast to the average correlation functions, the typical correlation functions:
$G_{\rm typ.}(r,r') = \exp\[\overline{\log\(\<\sigma^z_r\sigma^z_{r'}\>\)}\]$ are not sensitive to rare configurations where $r,r'$ belong to the same cluster, and are instead controlled by quantum tunneling (enhanced by the infinite-randomness critical dynamics), and expected to behave as: $G_{\rm typ.}(r,r')\sim e^{-|r-r'|^\psi}$ where $\psi$ is a universal tunneling dynamics exponent. 
For a broad class of $1d$ infinite-randomness critical points, including the random Ising universality class, $\psi_{1d}=1/2$.~\cite{fisher1994random}
This tunneling exponent also controls the dynamical relation between distance, $l$, and characteristic energy, $E$, in the strong-randomness critical point:
\begin{align}
    \log E(l)\sim l^\psi.
    \label{eq:psi}
\end{align}

For bulk criticality (with periodic boundary conditions)
the latter relation suggests the following procedure to determine $\psi$ within the strong-disorder RG approach: starting in a system with linear size $L$, the RG procedure is carried out until all the spins in the system are decimated. The (logarithmic) energy scale of the last decimation $
\log E(L)$ is recorded and averaged over many disorder instances. Repeating this procedure for a range of system sizes $L$, and fitting the resulting $\log E(L)$ versus $L$ provides an estimate: 
\begin{align}
    \psi_{\rm bulk}=0.50\pm 0.03.
\end{align}
This is consistent with the value of $0.48(2)$ obtained in Ref.~\cite{kovacs2010renormalization}, and slightly larger than that, ($\psi_{\rm bulk}=)0.42\pm 0.06$) reported in the earlier work Ref.~\cite{motrunich2000infinite} (which we speculate may arise from the less reliable fitting procedure used to locate $w_c$ in that work).
%By contrast Ref.~\cite{motrunich2000infinite} reported a slightly smaller value $\psi_{\rm bulk}=0.42\pm 0.06$ (though the error bars in these estimates do, barely, overlap). We suspect that the discrepancy here arises dues to two reasons. The first is the value of $w$ at criticality. Our analysis of the percolation and the bulk magnetization determines the critical $w$ to be $w=6.61$, which is slightly different from Motrunich et al. ($w=6.67$). In fact, using a larger value of $w$ lowers the fitted value of the $\psi$ exponent. The second reason is the different method used to extract the exponent. Motrunich et al. extract it directly from the slope of the bond distribution at criticality, while we use the scaling of the logarithmic energy scale. We employ this alternate method since it is less sensitive to fitting errors, and is easily adapted to study boundary criticality. We note that \gaurav{Ref.~\cite{kovacs2010renormalization} performed an improved calculation for the 2D bulk SD fixed point, and reports a value of 0.48(2), which agrees more closely with our value of $\psi$.}

\paragraph{Fractal dimension of critical FM clusters}
The magnetic moment of clusters that can be magnetized with a field of strength $\sim E(\ell)$ scales as:
\begin{align}
    \mu \sim (\log E(l))^\phi \sim l^{d_f},
    \label{eq:phi}
\end{align}
where, using Eq.~\ref{eq:psi},
\begin{align} 
d_f=\psi\phi, \end{align} 
is the fractal dimension of the cluster.
%Combing with Eq. \ref{eq:psi}, this gives us an expression for the fractal dimension of the system:
%\begin{align}
%    \mu \sim l^{d_f} \\
%    d_f = \psi \phi
%\end{align}
We can numerically extract $d_f$ by fitting the average moment of the system, given in the RG by the number of spins in undecimated FM clusters at a given RG step, to a power law. From this procedure, we find \begin{align}
    d_f=0.98\pm0.05 ~~~~{\rm (bulk)}
\end{align}
which matches the value determined by \cite{motrunich2000infinite} $d_f=1.0\pm0.1$.

\section{Boundary Criticality}
\begin{figure}[t]
%\begin{center}
\includegraphics[width=0.45\textwidth]{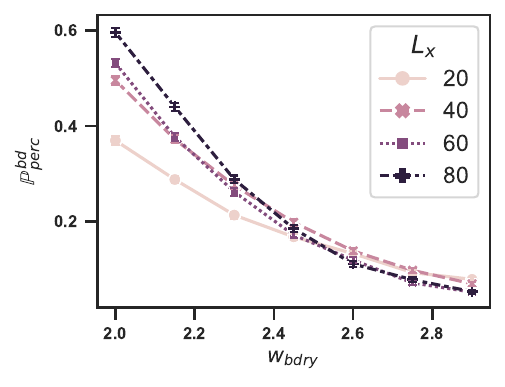}
%{\Large $(b)$}\includegraphics[width=0.45\textwidth]{images/OBC_magnetization_ratio.pdf}
%\end{center}
\caption{{\bf Boundary criticality.} 
 Probability to find a ferromagnetic cluster percolating around the top boundary, as a function of the width, $w_{\rm bdry}$ of the boundary-boundary bond distribution. In a similar fashion as before, a system is marked as percolating if two spins in zones of given radius at $x=L$ and $x=L_x/2$ belong to the same cluster. The boundary size $L_x$ is varied while the bulk size is fixed at $L_y=40$. The data is averaged over $1000$ disorder instances. We note that the decay of the percolation probability in the ordinary phase is slower than exponential due algebraic (average) bulk correlations.}
\label{fig:bdry_criticality}
\end{figure}

\begin{figure*}[t]
\begin{overpic}[width=0.7\textwidth]{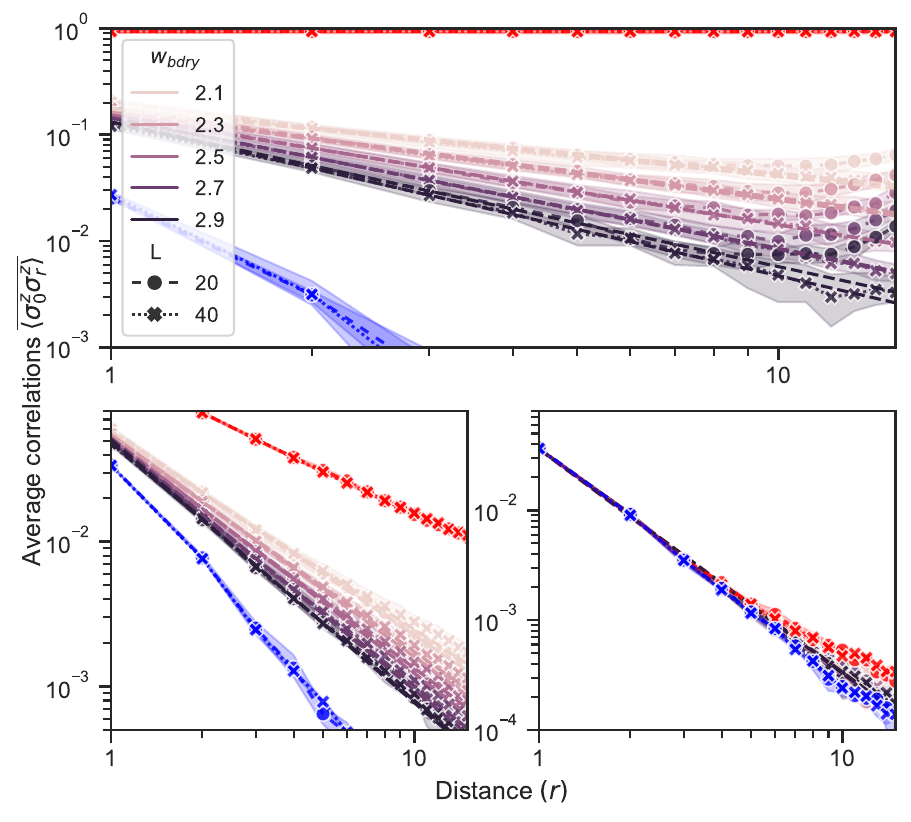}
\put(80,80){\Large $(a)$}
\put(20,15){\Large $(b)$}
\put(65,15){\Large $(c)$}
\end{overpic}
\caption{{\bf Average spin correlations.} Averaged correlations for systems of boundary size $L_x=L$ and fixed bulk size $L_y=50$. The bulk is critical while the width of the boundary-boundary bond distribution $w_{\rm bdry}$ is being varied. Panel a) depicts the boundary-boundary correlation functions. Panel b) depicts the boundary-bulk correlation functions. Panel c) depicts the bulk-bulk correlation functions. Solid lines are power law fits to the data, with the exponent denoted as $\phi$. The red and blue lines correspond to points deep in the extra-ordinary and ordinary phases respectively, corresponding to $w_{\rm bdry}=0.1$ and $7.0$ respectively.
}
\label{fig:corr_func}
\end{figure*}

To investigate boundary criticality, we switch from periodic boundary conditions, to a cylindrical geometry that is open along the $y$ direction with length $L_y$ and periodic in the $x$ direction with circumference $L_x$. 
As the RG proceeds, we label any spin cluster containing a spin originally at the boundary of the cylinder as a \emph{boundary cluster}, and those that contain only bulk spins as \emph{bulk clusters}.
%As the RG proceeds, we identify any spin cluster that contains a spin that was originally on the boundary of the cylinder as a boundary cluster. 
%
This results in three distinct types (BB,BS,SS) of exchange coupling distributions for bonds connecting bulk (B) or boundary/surface (S) clusters interactions.
Throughout the rest of the discussion, we fix the bulk coupling distributions at the previously determined critical point and tune only the boundary-boundary bond couplings to determine the boundary criticality classes. 
In order to approximately isolate a single boundary, we only tune the coupling distribution on the top boundary, $w_{\rm bdry}$, and leave the bottom boundary bonds at $w_{\rm bottom-bdry}=6.6\approx w_{c,{\rm bulk}}$, which we find is deep in the ordinary boundary critical phase (and hence does not induce a large boundary-magnetization that would span the bulk in a finite-size system and contaminate the top boundary).

\subsection{Boundary critical phase diagram}
As for the clean Ising model, we observe three distinct boundary critical classes shown in (Fig.~\ref{fig:phase_diag}): 
\begin{enumerate}
    \item Ordinary or boundary/surface-disordered (SD): At large $w_{\rm bdry}$, the boundary transverse fields dominate at the boundary such that by itself, the boundary would be deep in the disordered paramagnetic phase, and the critical behavior of the boundary spins is ``inherited" from the bulk critical fluctuations.
    \item Extraordinary or boundary/surface-ordered (SO): at small $w_{\rm bdry}$, the ferromagnetic exchange couplings at the boundary dominate over the transverse fields and give rise to long-range symmetry-breaking order on the boundary, signaled by the emergence of a single ferromagnetic cluster spanning the boundary. The magnetization of this cluster decays slowly into the critically bulk which has large spin susceptibility.
    \item Special: Finally, there is a boundary phase transition at $w_{\rm bdry}=w_{\rm sp}$ separating the above two boundary critical phases, which does not have long-range symmetry-breaking order at the boundary, but exhibits distinct boundary criticality from the ordinary phase due to the interplay of bulk and boundary critical fluctuations
\end{enumerate}
As for the bulk critical point, the special transition can be located 
%either i) 
via the finite-size crossing in the probability, $\mathbb{P}_{\rm perc,bdry}$, to find a ferromagnetic cluster percolating around the boundary.
%or ii) via the ratio, $m_{\rm bdry}(2L)/m_{\rm bdry}(L)$ of boundary magnetizations (Fig.~\ref{fig:bdry_criticality}). We define boundary versions of both of these quantities in the same manner as for the bulk transition discussed above, but restricted to boundary clusters.
The crossing occurs at critical value of boundary coupling widths, 
\begin{align}
w_{c,{\rm bdry}}=w_{sp}\approx 2.4.
\end{align}
%with crossing value, 
%$m_{\rm bdry}(2L)/m_{\rm bdry}(L)\vert_{w_{\rm bdry}=w_{sp}} \approx  0.39\pm 0.1$.

In passing, we note a practical complication for finite-size analysis. Due to the bulk criticality, 
$\mathbb{P}_{\rm perc,bdry}$ 
decays only algebraically in the ordinary boundary critical phase (compared to the exponential decay of their bulk analogs in the paramagnetic phase). This exacerbates finite-size drifts in the crossing value.

\subsection{Correlation Functions and Dynamics}
The different boundary critical classes exhibit markedly different behaviors in two-point spin-correlations:
between boundary spins, and between bulk- and boundary-spins.

\paragraph{Average correlation functions}
With open boundary conditions, the asymptotic long-distance behavior of $\overline{G}(r,r')$ depends on the boundary universality class and whether $r$ and $r'$ belong to the bulk (B) or boundary/surface (S). 
In all cases but one, the correlations decay as a power-law with the distance $|r-r'|$: \begin{align}
    \overline{G(r,r')} \sim |r-r'|^{-\eta},
    \label{eq:eta}
\end{align}
where the universal exponents $\eta$ for various boundary universality classes and locations of points $r,r'$ are listed in Table.~\ref{tab:eta_exps}.
The sole exception to (\ref{eq:eta}) is for extraordinary boundary criticality class, for which the boundary/boundary (SS) correlations exhibit non-decaying long-range order (LRO), corresponding to $\eta=0$.

While the bulk correlations decay with an inverse square of the distance in all cases, at the ordinary (boundary-disordered) critical point, the range of correlations involving one or more boundary (S) spin is suppressed (larger $\eta$) compared to the bulk due to the enhanced transverse fields near the boundary that suppress FM order near the sample's edge.
Conversely, at the special transition where the boundary is on the verge of developing long-range order, the range of FM correlations involving boundary spins is enhanced (smaller $\eta$). 

More precisely, note that $\eta_{\rm BB} = 2 x_m$ where $x_m$ is the (bulk) scaling dimension of the magnetization operator in eq.~\eqref{eqRatiom}, so $\eta_{\rm BB} \simeq 2.0$ since $x_m \simeq 1.0$. The boundary values of $\eta$ are then given by $\eta_{\rm SB} = x_m + x_{m_{\rm bdry}}$ and $\eta_{\rm SS} = 2 x_{m_{\rm bdry}}$ where $x_{m_{\rm bdry}}$ is the {\rm boundary} magnetization scaling dimension. In the extraordinary case, we have $x_{m_{\rm bdry}}=0$ since the boundary is ordered, so that $\eta_{\rm SS} = 0 $ and $\eta_{\rm SB} \simeq 1.0 $. In the ordinary and special cases, our results in Tab.~\ref{tab:eta_exps} indicate that   $x_{m_{\rm bdry}}\simeq 1.5 $ and $x_{m_{\rm bdry}}=0.5$, respectively.

\paragraph{Tunneling dynamical exponent, $\psi$:}
\begin{figure}
%\begin{center}
{\Large $(a)$}\includegraphics[width=0.4\textwidth]{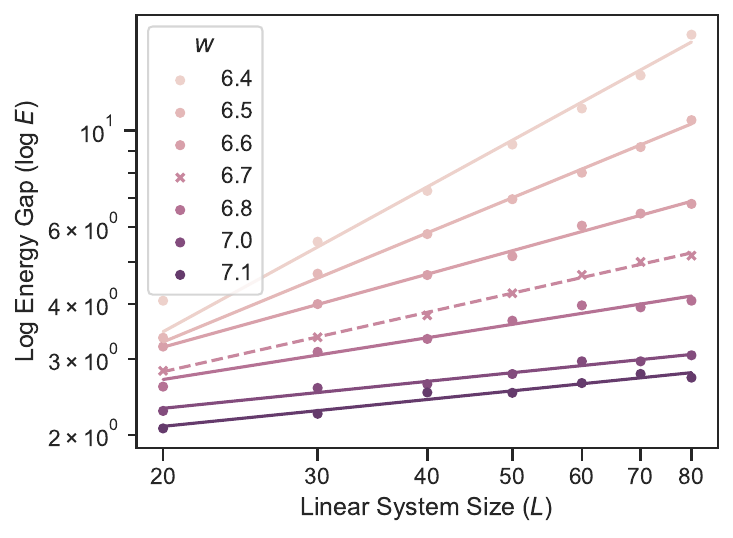}
{\Large $(b)$}\includegraphics[width=0.4\textwidth]{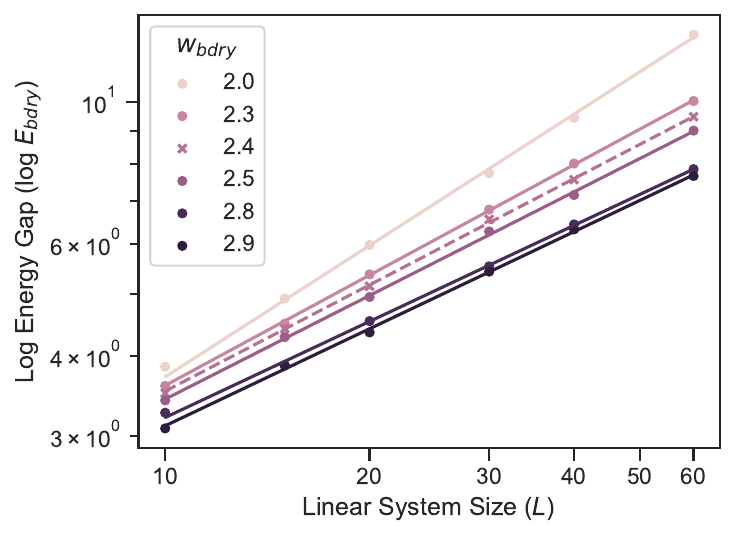}
%\end{center}
\caption{{\bf Gap scaling.} (a) Scaling of log $E$ for the PBC system versus the linear system size. The dashed line corresponds to $w=6.7$. The $\psi$ exponent is obtained by a power law fit at the critical point, with error determined by the uncertainty in the critical parameter. (b) Scaling of log $E_{\rm bdry}$ versus the linear system size. The scaling at the special critical point ($w_{\rm bdry}=2.4$) is indicated by a dashed line. Observe that in the ordinary regime ($w_{\rm bdry}>2.4$), the gap scaling approaches the 1D RFIM limit of $\psi=0.5$}
\label{fig:bdry_criticality}
\end{figure}
The $\psi$ exponent can be extracted by a simple modification of the above-described procedure for $\psi_{\rm bulk}$: Starting with a system of linear size $L$ and open boundary conditions, the RG procedure is iterated until one entire boundary has been merged into a single cluster. The logarithmic energy scale, $\log E_{\rm bdry}(L)$, is recorded for the final boundary decimation.
Fitting the resulting $\psi$ exponent (Table~\ref{tab:psi_exps}), we observe that for both the special and ordinary transition $\psi_{\rm bdry}$ is quite close to the standard $1d$ value of $0.5$.
By contrast, for the extraordinary (boundary-ordered) boundary critical class, the boundary exhibits LRO, i.e. the bondary spins are merged into a single ferromagnetic cluster at some $O(1)$ energy scale independent of system size $L$, and the scaling form of (\ref{eq:psi}) does not apply.

\begin{table}[t!]
\renewcommand{\arraystretch}{1.3}
    \centering
    \begin{tabular}{|l|c|c|c|}
        \hline
        Boundary Class & $\eta_{\rm BB}$ & $\eta_{\rm SB}$ & $\eta_{SS}$ \\ 
        \hline
        Ordinary & $2.05\pm0.01$ & $2.51\pm0.04$ & $3.02\pm0.03$ \\
        %\hline
        Extraordinary & $2.02\pm 0.01$ & $1.01\pm 0.01$ & 0 (LRO) \\ 
        %\hline
        Special & $2.04\pm0.01$ & $1.51\pm0.07$ & $0.9\pm0.1$ \\ 
         \hline
    \end{tabular}
    \caption{$\eta$ exponent derived from power law fitting the average correlation functions with sites $r$ and $r'$ in the bulk (B) or surface/boundary (S).
    LRO indicates long-range order ($\overline{G}$ asymptotes to a non-vanishing constant, i.e. $\eta=0$).
    }
    \label{tab:eta_exps}
\end{table}

\begin{table}[t!]
\renewcommand{\arraystretch}{1.3}
    \centering
    \begin{tabular}{|l|c|}
        \hline
        Boundary Class& $\psi_{\rm bdry}$ \\ \hline
         %\hline
        Ordinary & $0.51\pm0.01 $ \\
        Extraordinary & 0 (LRO) \\ 
        Special & $0.54\pm0.01$\\
        \hline
    \end{tabular}
    \caption{Boundary dynamical tunneling critical exponent $\psi$ relating the scalign of energy $E$ and distance $L$ as $\log E(L)\sim -L^\psi$.}
    \label{tab:psi_exps}
\end{table}

\section{Discussion and Outlook}
In this work, we have numerically studied the boundary criticality of a random $2d$ transverse field Ising model at strong disorder.
This study establishes the strong-disorder RG method as a useful numerical tool for studying boundary quantum criticality in systems that are far too large to treat by direct classical simulation.
One intriguing observation is that many of the scaling exponents observed (both bulk and boundary) have values very close to rational values (see Tables~\ref{tab:psi_exps},\ref{tab:psi_exps}) -- a rarity for phase transitions in more than one spatial dimension.
In particular numerical value of the tunneling exponent $\psi$ is consistent with $1/2$ independent of dimensionality or bulk versus boundary criticality. 
It would be interesting to explore whether the rationality of these exponents could be established directly by an analytic means.

We anticipate that this approach can be readily adapted to explore a large class disordered boundary critical phenomena.
A particularly interesting target would be symmetry- and topology-enriched critical phenomena in systems with topological edge states, such as gapless symmetry-protected states (gSPTs)~\cite{scaffidi2017gapless,verresen2021gapless} (see~\cite{PhysRevB.103.L100207} for a one-dimensional example of symmetry-enriched infinite randomness critical point).
Another natural target of interest would be disordered 2d states with intrinsic topological order~\cite{PhysRevB.102.224204}. 
For example, while the bulk Higgs and confined phases of discrete gauge theories are smoothly connected without a phase transition, recent work shows that they are distinguished by higher-form SPT invariants and are distinguished by a boundary phase transition~\cite{verresen2022higgs}. The strong-disorder RG procedure offers a promising numerical tool to study disordered discrete lattice gauge theory models and explore these issues with a well-controlled numerical tools.

\paragraph*{Acknowledgments -- }
We thank Alberto Nocera for insightful conversations and Sid Parameswaran and Snir Gazit for previous collaborations on related topics.
This research was supported by the Natural Sciences and Engineering Research Council of Canada (NSERC), by the US Department of Energy, Office of Science, Basic Energy Sciences, under award No. DE-SC0023999 (R.V.), and in part through computational resources and services provided by Advanced Research Computing at the University of British Columbia.
This work was performed in part at the Aspen Center for Physics, which is supported by National Science Foundation grant PHY-2210452.

\bibliography{rsrg_bib}% Produces the bibliography via BibTeX.

\begin{thebibliography}{24}%
\makeatletter
\providecommand \@ifxundefined [1]{%
 \@ifx{#1\undefined}
}%
\providecommand \@ifnum [1]{%
 \ifnum #1\expandafter \@firstoftwo
 \else \expandafter \@secondoftwo
 \fi
}%
\providecommand \@ifx [1]{%
 \ifx #1\expandafter \@firstoftwo
 \else \expandafter \@secondoftwo
 \fi
}%
\providecommand \natexlab [1]{#1}%
\providecommand \enquote  [1]{``#1''}%
\providecommand \bibnamefont  [1]{#1}%
\providecommand \bibfnamefont [1]{#1}%
\providecommand \citenamefont [1]{#1}%
\providecommand \href@noop [0]{\@secondoftwo}%
\providecommand \href [0]{\begingroup \@sanitize@url \@href}%
\providecommand \@href[1]{\@@startlink{#1}\@@href}%
\providecommand \@@href[1]{\endgroup#1\@@endlink}%
\providecommand \@sanitize@url [0]{\catcode `\\12\catcode `\$12\catcode
  `\&12\catcode `\#12\catcode `\^12\catcode `\_12\catcode `\%12\relax}%
\providecommand \@@startlink[1]{}%
\providecommand \@@endlink[0]{}%
\providecommand \url  [0]{\begingroup\@sanitize@url \@url }%
\providecommand \@url [1]{\endgroup\@href {#1}{\urlprefix }}%
\providecommand \urlprefix  [0]{URL }%
\providecommand \Eprint [0]{\href }%
\providecommand \doibase [0]{http://dx.doi.org/}%
\providecommand \selectlanguage [0]{\@gobble}%
\providecommand \bibinfo  [0]{\@secondoftwo}%
\providecommand \bibfield  [0]{\@secondoftwo}%
\providecommand \translation [1]{[#1]}%
\providecommand \BibitemOpen [0]{}%
\providecommand \bibitemStop [0]{}%
\providecommand \bibitemNoStop [0]{.\EOS\space}%
\providecommand \EOS [0]{\spacefactor3000\relax}%
\providecommand \BibitemShut  [1]{\csname bibitem#1\endcsname}%
\let\auto@bib@innerbib\@empty
%</preamble>
\bibitem [{\citenamefont {Diehl}\ and\ \citenamefont
  {Lam}(1989)}]{diehl1989semiinfinite}%
  \BibitemOpen
  \bibfield  {author} {\bibinfo {author} {\bibfnamefont {H.~W.}\ \bibnamefont
  {Diehl}}\ and\ \bibinfo {author} {\bibfnamefont {P.~M.}\ \bibnamefont
  {Lam}},\ }\bibfield  {title} {\enquote {\bibinfo {title} {Semi-infinite potts
  model and percolation at surfaces},}\ }\href {\doibase 10.1007/bf01307889}
  {\bibfield  {journal} {\bibinfo  {journal} {Zeitschrift fur Physik B
  Condensed Matter}\ }\textbf {\bibinfo {volume} {74}},\ \bibinfo {pages}
  {395–401} (\bibinfo {year} {1989})}\BibitemShut {NoStop}%
\bibitem [{\citenamefont {Deng}\ \emph {et~al.}(2005)\citenamefont {Deng},
  \citenamefont {Bl\"ote},\ and\ \citenamefont
  {Nightingale}}]{deng2005surface}%
  \BibitemOpen
  \bibfield  {author} {\bibinfo {author} {\bibfnamefont {Youjin}\ \bibnamefont
  {Deng}}, \bibinfo {author} {\bibfnamefont {Henk W.~J.}\ \bibnamefont
  {Bl\"ote}}, \ and\ \bibinfo {author} {\bibfnamefont {M.~P.}\ \bibnamefont
  {Nightingale}},\ }\bibfield  {title} {\enquote {\bibinfo {title} {Surface and
  bulk transitions in three-dimensional $\mathrm{O}(n)$ models},}\ }\href
  {\doibase 10.1103/PhysRevE.72.016128} {\bibfield  {journal} {\bibinfo
  {journal} {Phys. Rev. E}\ }\textbf {\bibinfo {volume} {72}},\ \bibinfo
  {pages} {016128} (\bibinfo {year} {2005})}\BibitemShut {NoStop}%
\bibitem [{\citenamefont {Deng}(2006)}]{deng2006bulk}%
  \BibitemOpen
  \bibfield  {author} {\bibinfo {author} {\bibfnamefont {Youjin}\ \bibnamefont
  {Deng}},\ }\bibfield  {title} {\enquote {\bibinfo {title} {Bulk and surface
  phase transitions in the three-dimensional $o(4)$ spin model},}\ }\href
  {\doibase 10.1103/PhysRevE.73.056116} {\bibfield  {journal} {\bibinfo
  {journal} {Phys. Rev. E}\ }\textbf {\bibinfo {volume} {73}},\ \bibinfo
  {pages} {056116} (\bibinfo {year} {2006})}\BibitemShut {NoStop}%
\bibitem [{\citenamefont {Ohno}\ and\ \citenamefont {Okabe}(1984)}]{ohno19841}%
  \BibitemOpen
  \bibfield  {author} {\bibinfo {author} {\bibfnamefont {Kaoru}\ \bibnamefont
  {Ohno}}\ and\ \bibinfo {author} {\bibfnamefont {Yutaka}\ \bibnamefont
  {Okabe}},\ }\bibfield  {title} {\enquote {\bibinfo {title} {The 1/n expansion
  for the extraordinary transition of semi-infinite system},}\ }\href@noop {}
  {\bibfield  {journal} {\bibinfo  {journal} {Progress of theoretical physics}\
  }\textbf {\bibinfo {volume} {72}},\ \bibinfo {pages} {736--745} (\bibinfo
  {year} {1984})}\BibitemShut {NoStop}%
\bibitem [{\citenamefont {Landau}\ \emph {et~al.}(1989)\citenamefont {Landau},
  \citenamefont {Pandey},\ and\ \citenamefont {Binder}}]{landau1989monte}%
  \BibitemOpen
  \bibfield  {author} {\bibinfo {author} {\bibfnamefont {D.~P.}\ \bibnamefont
  {Landau}}, \bibinfo {author} {\bibfnamefont {R.}~\bibnamefont {Pandey}}, \
  and\ \bibinfo {author} {\bibfnamefont {K.}~\bibnamefont {Binder}},\
  }\bibfield  {title} {\enquote {\bibinfo {title} {Monte carlo study of surface
  critical behavior in the xy model},}\ }\href {\doibase
  10.1103/PhysRevB.39.12302} {\bibfield  {journal} {\bibinfo  {journal} {Phys.
  Rev. B}\ }\textbf {\bibinfo {volume} {39}},\ \bibinfo {pages} {12302--12305}
  (\bibinfo {year} {1989})}\BibitemShut {NoStop}%
\bibitem [{\citenamefont {Diehl}(1997)}]{diehl1997thetheory}%
  \BibitemOpen
  \bibfield  {author} {\bibinfo {author} {\bibfnamefont {H.~W.}\ \bibnamefont
  {Diehl}},\ }\bibfield  {title} {\enquote {\bibinfo {title} {The theory of
  boundary critical phenomena},}\ }\href {\doibase 10.1142/S0217979297001751}
  {\bibfield  {journal} {\bibinfo  {journal} {International Journal of Modern
  Physics B}\ }\textbf {\bibinfo {volume} {11}},\ \bibinfo {pages} {3503--3523}
  (\bibinfo {year} {1997})},\ \Eprint
  {http://arxiv.org/abs/https://doi.org/10.1142/S0217979297001751}
  {https://doi.org/10.1142/S0217979297001751} \BibitemShut {NoStop}%
\bibitem [{\citenamefont {Cardy}(1996)}]{cardy1996scaling}%
  \BibitemOpen
  \bibfield  {author} {\bibinfo {author} {\bibfnamefont {John}\ \bibnamefont
  {Cardy}},\ }\href@noop {} {\emph {\bibinfo {title} {Scaling and
  Renormalization in Statistical Physics}}},\ Cambridge Lecture Notes in
  Physics\ (\bibinfo  {publisher} {Cambridge University Press},\ \bibinfo
  {year} {1996})\BibitemShut {NoStop}%
\bibitem [{\citenamefont {Metlitski}(2022)}]{metlitski2022on}%
  \BibitemOpen
  \bibfield  {author} {\bibinfo {author} {\bibfnamefont {Max~A.}\ \bibnamefont
  {Metlitski}},\ }\bibfield  {title} {\enquote {\bibinfo {title} {{Boundary
  criticality of the O(N) model in d = 3 critically revisited}},}\ }\href
  {\doibase 10.21468/SciPostPhys.12.4.131} {\bibfield  {journal} {\bibinfo
  {journal} {SciPost Phys.}\ }\textbf {\bibinfo {volume} {12}},\ \bibinfo
  {pages} {131} (\bibinfo {year} {2022})}\BibitemShut {NoStop}%
\bibitem [{\citenamefont {Sun}\ and\ \citenamefont
  {Lv}(2022)}]{sun2022quantum}%
  \BibitemOpen
  \bibfield  {author} {\bibinfo {author} {\bibfnamefont {Yanan}\ \bibnamefont
  {Sun}}\ and\ \bibinfo {author} {\bibfnamefont {Jian-Ping}\ \bibnamefont
  {Lv}},\ }\bibfield  {title} {\enquote {\bibinfo {title} {Quantum
  extraordinary-log universality of boundary critical behavior},}\ }\href
  {\doibase 10.1103/PhysRevB.106.224502} {\bibfield  {journal} {\bibinfo
  {journal} {Phys. Rev. B}\ }\textbf {\bibinfo {volume} {106}},\ \bibinfo
  {pages} {224502} (\bibinfo {year} {2022})}\BibitemShut {NoStop}%
\bibitem [{\citenamefont {Weber}\ \emph {et~al.}(2018)\citenamefont {Weber},
  \citenamefont {Parisen~Toldin},\ and\ \citenamefont
  {Wessel}}]{weber2018nonordinary}%
  \BibitemOpen
  \bibfield  {author} {\bibinfo {author} {\bibfnamefont {Lukas}\ \bibnamefont
  {Weber}}, \bibinfo {author} {\bibfnamefont {Francesco}\ \bibnamefont
  {Parisen~Toldin}}, \ and\ \bibinfo {author} {\bibfnamefont {Stefan}\
  \bibnamefont {Wessel}},\ }\bibfield  {title} {\enquote {\bibinfo {title}
  {Nonordinary edge criticality of two-dimensional quantum critical magnets},}\
  }\href {\doibase 10.1103/PhysRevB.98.140403} {\bibfield  {journal} {\bibinfo
  {journal} {Phys. Rev. B}\ }\textbf {\bibinfo {volume} {98}},\ \bibinfo
  {pages} {140403} (\bibinfo {year} {2018})}\BibitemShut {NoStop}%
\bibitem [{\citenamefont {Verresen}\ \emph {et~al.}(2022)\citenamefont
  {Verresen}, \citenamefont {Borla}, \citenamefont {Vishwanath}, \citenamefont
  {Moroz},\ and\ \citenamefont {Thorngren}}]{verresen2022higgs}%
  \BibitemOpen
  \bibfield  {author} {\bibinfo {author} {\bibfnamefont {Ruben}\ \bibnamefont
  {Verresen}}, \bibinfo {author} {\bibfnamefont {Umberto}\ \bibnamefont
  {Borla}}, \bibinfo {author} {\bibfnamefont {Ashvin}\ \bibnamefont
  {Vishwanath}}, \bibinfo {author} {\bibfnamefont {Sergej}\ \bibnamefont
  {Moroz}}, \ and\ \bibinfo {author} {\bibfnamefont {Ryan}\ \bibnamefont
  {Thorngren}},\ }\bibfield  {title} {\enquote {\bibinfo {title} {Higgs
  condensates are symmetry-protected topological phases: I. discrete
  symmetries},}\ }\href@noop {} {\bibfield  {journal} {\bibinfo  {journal}
  {arXiv preprint arXiv:2211.01376}\ } (\bibinfo {year} {2022})}\BibitemShut
  {NoStop}%
\bibitem [{\citenamefont {Dasgupta}\ and\ \citenamefont
  {Ma}(1980)}]{ma1980lowtemperature}%
  \BibitemOpen
  \bibfield  {author} {\bibinfo {author} {\bibfnamefont {Chandan}\ \bibnamefont
  {Dasgupta}}\ and\ \bibinfo {author} {\bibfnamefont {Shang-keng}\ \bibnamefont
  {Ma}},\ }\bibfield  {title} {\enquote {\bibinfo {title} {Low-temperature
  properties of the random heisenberg antiferromagnetic chain},}\ }\href
  {\doibase 10.1103/PhysRevB.22.1305} {\bibfield  {journal} {\bibinfo
  {journal} {Phys. Rev. B}\ }\textbf {\bibinfo {volume} {22}},\ \bibinfo
  {pages} {1305--1319} (\bibinfo {year} {1980})}\BibitemShut {NoStop}%
\bibitem [{\citenamefont {Fisher}(1992)}]{fisher1992random}%
  \BibitemOpen
  \bibfield  {author} {\bibinfo {author} {\bibfnamefont {Daniel~S.}\
  \bibnamefont {Fisher}},\ }\bibfield  {title} {\enquote {\bibinfo {title}
  {Random transverse field ising spin chains},}\ }\href {\doibase
  10.1103/PhysRevLett.69.534} {\bibfield  {journal} {\bibinfo  {journal} {Phys.
  Rev. Lett.}\ }\textbf {\bibinfo {volume} {69}},\ \bibinfo {pages} {534--537}
  (\bibinfo {year} {1992})}\BibitemShut {NoStop}%
\bibitem [{\citenamefont {Fisher}(1995)}]{fisher1995critical}%
  \BibitemOpen
  \bibfield  {author} {\bibinfo {author} {\bibfnamefont {Daniel~S.}\
  \bibnamefont {Fisher}},\ }\bibfield  {title} {\enquote {\bibinfo {title}
  {Critical behavior of random transverse-field ising spin chains},}\ }\href
  {\doibase 10.1103/PhysRevB.51.6411} {\bibfield  {journal} {\bibinfo
  {journal} {Phys. Rev. B}\ }\textbf {\bibinfo {volume} {51}},\ \bibinfo
  {pages} {6411--6461} (\bibinfo {year} {1995})}\BibitemShut {NoStop}%
\bibitem [{\citenamefont {Motrunich}\ \emph {et~al.}(2000)\citenamefont
  {Motrunich}, \citenamefont {Mau}, \citenamefont {Huse},\ and\ \citenamefont
  {Fisher}}]{motrunich2000infinite}%
  \BibitemOpen
  \bibfield  {author} {\bibinfo {author} {\bibfnamefont {Olexei}\ \bibnamefont
  {Motrunich}}, \bibinfo {author} {\bibfnamefont {Siun-Chuon}\ \bibnamefont
  {Mau}}, \bibinfo {author} {\bibfnamefont {David~A.}\ \bibnamefont {Huse}}, \
  and\ \bibinfo {author} {\bibfnamefont {Daniel~S.}\ \bibnamefont {Fisher}},\
  }\bibfield  {title} {\enquote {\bibinfo {title} {Infinite-randomness quantum
  ising critical fixed points},}\ }\href {\doibase 10.1103/PhysRevB.61.1160}
  {\bibfield  {journal} {\bibinfo  {journal} {Phys. Rev. B}\ }\textbf {\bibinfo
  {volume} {61}},\ \bibinfo {pages} {1160--1172} (\bibinfo {year}
  {2000})}\BibitemShut {NoStop}%
\bibitem [{\citenamefont {Pandey}\ \emph {et~al.}(2023)\citenamefont {Pandey},
  \citenamefont {Mahadevan},\ and\ \citenamefont {Cowsik}}]{pandey2023random}%
  \BibitemOpen
  \bibfield  {author} {\bibinfo {author} {\bibfnamefont {Akshat}\ \bibnamefont
  {Pandey}}, \bibinfo {author} {\bibfnamefont {Aditya}\ \bibnamefont
  {Mahadevan}}, \ and\ \bibinfo {author} {\bibfnamefont {Aditya}\ \bibnamefont
  {Cowsik}},\ }\bibfield  {title} {\enquote {\bibinfo {title} {Random geometry
  at an infinite-randomness fixed point},}\ }\href {\doibase
  10.1103/PhysRevB.108.064201} {\bibfield  {journal} {\bibinfo  {journal}
  {Phys. Rev. B}\ }\textbf {\bibinfo {volume} {108}},\ \bibinfo {pages}
  {064201} (\bibinfo {year} {2023})}\BibitemShut {NoStop}%
\bibitem [{\citenamefont {Kov\'acs}\ and\ \citenamefont
  {Igl\'oi}(2010)}]{kovacs2010renormalization}%
  \BibitemOpen
  \bibfield  {author} {\bibinfo {author} {\bibfnamefont {Istv\'an~A.}\
  \bibnamefont {Kov\'acs}}\ and\ \bibinfo {author} {\bibfnamefont {Ferenc}\
  \bibnamefont {Igl\'oi}},\ }\bibfield  {title} {\enquote {\bibinfo {title}
  {Renormalization group study of the two-dimensional random transverse-field
  ising model},}\ }\href {\doibase 10.1103/PhysRevB.82.054437} {\bibfield
  {journal} {\bibinfo  {journal} {Phys. Rev. B}\ }\textbf {\bibinfo {volume}
  {82}},\ \bibinfo {pages} {054437} (\bibinfo {year} {2010})}\BibitemShut
  {NoStop}%
\bibitem [{\citenamefont {Yu}\ \emph {et~al.}(2008)\citenamefont {Yu},
  \citenamefont {Saleur},\ and\ \citenamefont {Haas}}]{yu2008entanglement}%
  \BibitemOpen
  \bibfield  {author} {\bibinfo {author} {\bibfnamefont {Rong}\ \bibnamefont
  {Yu}}, \bibinfo {author} {\bibfnamefont {Hubert}\ \bibnamefont {Saleur}}, \
  and\ \bibinfo {author} {\bibfnamefont {Stephan}\ \bibnamefont {Haas}},\
  }\bibfield  {title} {\enquote {\bibinfo {title} {Entanglement entropy in the
  two-dimensional random transverse field ising model},}\ }\href {\doibase
  10.1103/PhysRevB.77.140402} {\bibfield  {journal} {\bibinfo  {journal} {Phys.
  Rev. B}\ }\textbf {\bibinfo {volume} {77}},\ \bibinfo {pages} {140402}
  (\bibinfo {year} {2008})}\BibitemShut {NoStop}%
\bibitem [{\citenamefont {Kov\'acs}\ and\ \citenamefont
  {Igl\'oi}(2013)}]{kovacs2013boundary}%
  \BibitemOpen
  \bibfield  {author} {\bibinfo {author} {\bibfnamefont {Istv\'an~A.}\
  \bibnamefont {Kov\'acs}}\ and\ \bibinfo {author} {\bibfnamefont {Ferenc}\
  \bibnamefont {Igl\'oi}},\ }\bibfield  {title} {\enquote {\bibinfo {title}
  {Boundary critical phenomena of the random transverse ising model in
  $d\ensuremath{\ge}2$ dimensions},}\ }\href {\doibase
  10.1103/PhysRevB.87.024204} {\bibfield  {journal} {\bibinfo  {journal} {Phys.
  Rev. B}\ }\textbf {\bibinfo {volume} {87}},\ \bibinfo {pages} {024204}
  (\bibinfo {year} {2013})}\BibitemShut {NoStop}%
\bibitem [{\citenamefont {Scaffidi}\ \emph {et~al.}(2017)\citenamefont
  {Scaffidi}, \citenamefont {Parker},\ and\ \citenamefont
  {Vasseur}}]{scaffidi2017gapless}%
  \BibitemOpen
  \bibfield  {author} {\bibinfo {author} {\bibfnamefont {Thomas}\ \bibnamefont
  {Scaffidi}}, \bibinfo {author} {\bibfnamefont {Daniel~E.}\ \bibnamefont
  {Parker}}, \ and\ \bibinfo {author} {\bibfnamefont {Romain}\ \bibnamefont
  {Vasseur}},\ }\bibfield  {title} {\enquote {\bibinfo {title} {Gapless
  symmetry-protected topological order},}\ }\href {\doibase
  10.1103/PhysRevX.7.041048} {\bibfield  {journal} {\bibinfo  {journal} {Phys.
  Rev. X}\ }\textbf {\bibinfo {volume} {7}},\ \bibinfo {pages} {041048}
  (\bibinfo {year} {2017})}\BibitemShut {NoStop}%
\bibitem [{\citenamefont {Verresen}\ \emph {et~al.}(2021)\citenamefont
  {Verresen}, \citenamefont {Thorngren}, \citenamefont {Jones},\ and\
  \citenamefont {Pollmann}}]{verresen2021gapless}%
  \BibitemOpen
  \bibfield  {author} {\bibinfo {author} {\bibfnamefont {Ruben}\ \bibnamefont
  {Verresen}}, \bibinfo {author} {\bibfnamefont {Ryan}\ \bibnamefont
  {Thorngren}}, \bibinfo {author} {\bibfnamefont {Nick~G.}\ \bibnamefont
  {Jones}}, \ and\ \bibinfo {author} {\bibfnamefont {Frank}\ \bibnamefont
  {Pollmann}},\ }\bibfield  {title} {\enquote {\bibinfo {title} {Gapless
  topological phases and symmetry-enriched quantum criticality},}\ }\href
  {\doibase 10.1103/PhysRevX.11.041059} {\bibfield  {journal} {\bibinfo
  {journal} {Phys. Rev. X}\ }\textbf {\bibinfo {volume} {11}},\ \bibinfo
  {pages} {041059} (\bibinfo {year} {2021})}\BibitemShut {NoStop}%
\bibitem [{\citenamefont {Fisher}(1994)}]{fisher1994random}%
  \BibitemOpen
  \bibfield  {author} {\bibinfo {author} {\bibfnamefont {Daniel~S.}\
  \bibnamefont {Fisher}},\ }\bibfield  {title} {\enquote {\bibinfo {title}
  {Random antiferromagnetic quantum spin chains},}\ }\href {\doibase
  10.1103/PhysRevB.50.3799} {\bibfield  {journal} {\bibinfo  {journal} {Phys.
  Rev. B}\ }\textbf {\bibinfo {volume} {50}},\ \bibinfo {pages} {3799--3821}
  (\bibinfo {year} {1994})}\BibitemShut {NoStop}%
\bibitem [{\citenamefont {Duque}\ \emph {et~al.}(2021)\citenamefont {Duque},
  \citenamefont {Hu}, \citenamefont {You}, \citenamefont {Khemani},
  \citenamefont {Verresen},\ and\ \citenamefont
  {Vasseur}}]{PhysRevB.103.L100207}%
  \BibitemOpen
  \bibfield  {author} {\bibinfo {author} {\bibfnamefont {Carlos~M.}\
  \bibnamefont {Duque}}, \bibinfo {author} {\bibfnamefont {Hong-Ye}\
  \bibnamefont {Hu}}, \bibinfo {author} {\bibfnamefont {Yi-Zhuang}\
  \bibnamefont {You}}, \bibinfo {author} {\bibfnamefont {Vedika}\ \bibnamefont
  {Khemani}}, \bibinfo {author} {\bibfnamefont {Ruben}\ \bibnamefont
  {Verresen}}, \ and\ \bibinfo {author} {\bibfnamefont {Romain}\ \bibnamefont
  {Vasseur}},\ }\bibfield  {title} {\enquote {\bibinfo {title} {Topological and
  symmetry-enriched random quantum critical points},}\ }\href {\doibase
  10.1103/PhysRevB.103.L100207} {\bibfield  {journal} {\bibinfo  {journal}
  {Phys. Rev. B}\ }\textbf {\bibinfo {volume} {103}},\ \bibinfo {pages}
  {L100207} (\bibinfo {year} {2021})}\BibitemShut {NoStop}%
\bibitem [{\citenamefont {Kang}\ \emph {et~al.}(2020)\citenamefont {Kang},
  \citenamefont {Parameswaran}, \citenamefont {Potter}, \citenamefont
  {Vasseur},\ and\ \citenamefont {Gazit}}]{PhysRevB.102.224204}%
  \BibitemOpen
  \bibfield  {author} {\bibinfo {author} {\bibfnamefont {Byungmin}\
  \bibnamefont {Kang}}, \bibinfo {author} {\bibfnamefont {S.~A.}\ \bibnamefont
  {Parameswaran}}, \bibinfo {author} {\bibfnamefont {Andrew~C.}\ \bibnamefont
  {Potter}}, \bibinfo {author} {\bibfnamefont {Romain}\ \bibnamefont
  {Vasseur}}, \ and\ \bibinfo {author} {\bibfnamefont {Snir}\ \bibnamefont
  {Gazit}},\ }\bibfield  {title} {\enquote {\bibinfo {title} {Superuniversality
  from disorder at two-dimensional topological phase transitions},}\ }\href
  {\doibase 10.1103/PhysRevB.102.224204} {\bibfield  {journal} {\bibinfo
  {journal} {Phys. Rev. B}\ }\textbf {\bibinfo {volume} {102}},\ \bibinfo
  {pages} {224204} (\bibinfo {year} {2020})}\BibitemShut {NoStop}%
\end{thebibliography}%

\end{document}